\documentclass[12pt]{iopart}

\usepackage{graphicx}
\usepackage{amssymb}
\usepackage{iopams}
\expandafter\let\csname equation*\endcsname\relax
\expandafter\let\csname endequation*\endcsname\relax
\usepackage{amsmath}

\begin{document}

\title[Spatial localization and pattern formation in discrete optomechanical cavities]{Spatial localization and pattern formation in discrete optomechanical cavities and arrays}

\author{J Ruiz-Rivas$^1$, G Patera$^2$, C Navarrete-Benlloch$^{3,4,5}$, E Rold\'an$^1$ and G J de Valc\'arcel$^1$}
\address{$^1$Departament d'\`Optica, Universitat de Val\`{e}ncia, Dr. Moliner 50, 46100--Burjassot, Spain}
\address{$^2$Univ. Lille, CNRS, UMR 8523 - PhLAM - Physique des Lasers Atomes et Mol\'{e}cules, F-59000 Lille, France}
\address{$^3$Wilczek Quantum Center, School of Physics and Astronomy, Shanghai Jiao Tong University, Shanghai 200240, China}
\address{$^4$Shanghai Research Center for Quantum Sciences, Shanghai 201315, China}
\address{$^5$Max-Planck-Institut f\"{u}r die Physik des Lichts, Staudtstrasse 2, 91058 Erlangen, Germany}

\begin{abstract}
We investigate theoretically the generation of nonlinear dissipative structures in optomechanical (OM) systems containing discrete arrays of mechanical resonators. We consider both hybrid models in which the optical system is a continuous multimode field, as it would happen in an OM cavity
containing an array of micro-mirrors, and also fully discrete models in which each mechanical resonator interacts with a single optical mode, making contact with Ludwig \& Marquardt [Phys. Rev. Lett. \textbf{101},
073603 (2013)]. Also, we study the connections between both types of models and continuous OM models. While all three types of models merge naturally in the limit of a large number of densely distributed mechanical resonators, we show that the spatial localization and the pattern formation found in continuous OM models can be still observed for a small number of mechanical elements, even in the
presence of finite-size effects, which we discuss. This opens new venues for
experimental approaches to the subject.
\end{abstract}

\noindent{\it keywords: cavity optomechanics, optomechanical arrays, photon and phonon localization, spatial solitons\/}
\maketitle

\section{Introduction}
The emergence of patterns that spontaneously break some spatial symmetry is widespread in nonlinear optical systems, 
especially in large aspect-ratio cavities. Such patterns show up across the plane transverse to the light propagation 
direction ---hence the name ``transverse patterns'' to refer to them--- and have been theoretically and experimentally 
investigated in many different nonlinear optical cavities~\cite{StaliunasVSM,Mandel05}. 
Their study constitutes a well developed discipline in modern nonlinear optics. In fact, these nonlinear patterns belong 
to the wider class of dissipative structures, which are structures that self-sustain out of thermal equilibrium by continuous interchange of energy with the environment~\cite{CH}. Part of the interest of this research program in nonlinear optics, 
apart from its intrinsic physical relevance, lies in the potential for optical information storage and processing of a particular
type of pattern, namely cavity solitons, which are localized structures that can be individually written, erased, 
and even moved without affecting neighbouring structures~\cite{Firth,Barland,Esteban2005,ackemann}.

Naturally, most of the studies on optical transverse patterns so far have considered cavities containing usual passive or active nonlinear materials, like two-level systems, Kerr media, or second-order nonlinear crystals, to cite a few. More recently, optomechanical (OM) systems have begun to be considered in this respect. In~\cite{Ruiz15} we theoretically analyzed the possibility of using the nonlinear coupling between the cavity field and a deformable mechanical element to generate transverse patterns in OM
cavities. These are conceptually simple systems, consisting of an optical resonator with mechanical degrees of freedom that
couple to the light oscillating inside it~\cite{Aspelmeyer}. The coupling appears either through radiation pressure (e.g., when the mechanical degree of freedom corresponds to the oscillation of a perfectly reflecting cavity mirror) or through dispersive effects 
(e.g., when the mechanical degrees of freedom correspond to the local displacement of a partially transmitting membrane).
These systems are receiving intense and continued attention mainly in the context of modern quantum technologies, where phenomena such as cooling~\cite{Arcizet2006,Gigan2006,Kleckner2006}, induced transparency~\cite{Weis2010,Safavi2011}, squeezing~\cite{Brook2012,Safavi2013,Purdy2013} as well as single-photon strong-coupling regime~\cite{Akram2010,Nunnenkamp2011,Verhagen2012} have been demonstrated over the last decade. Generating dissipative structures in OM cavities could endow them with new capabilities. 
It could also lead to a pattern forming system in which quantum fluctuations play an important role: as OM cavities have demonstrated their ability to work within the quantum regime, they open the possibility of studying quantum dissipative structures under
the strong influence of quantum fluctuations. In the past some exciting phenomena concerning quantum fluctuations in dissipative structures were predicted for optical parametric oscillators~\cite{LuG,LuGatti,Santagiustina,Perez1,Perez2,Navarrete08}, however those models are far from realistic experimental implementations~\cite{Fabre}. Contrarily, OM cavities have demonstrated the feasibility of the simplest models as well as a large versatility because of the variety of possible platforms, materials and designs~\cite{Aspelmeyer,Safavi19,Liu19}. As a first step towards understanding this fully quantum picture, it is important to characterize
the conditions required for pattern formation in OM cavities.

Within the context of extended OM systems, Rakich and Marquardt \cite{Rakich18} have recently formulated a quantum theory of continuum optomechanics that is closely related to our theory~\cite{Ruiz15}. The theory is intended for treating OM interactions occurring along extended waveguides, which are naturally space-dependent problems in which Brillouin scattering is the essential coupling, thus connecting Brillouin physics with optomechanics. Below we address the formal connection between continuum
optomechanics and pattern forming OM cavities.

In \cite{Ruiz15} we demonstrated the feasibility of pattern formation in OM cavities by modeling the mecanical element, be it an end mirror or an intracavity membrane, as a continuously deformable element. Interestingly, a certain necessary condition, alien to other nonlinear optical cavities, must be fulfilled for nonlinear pattern formation in OM cavities, namely that the mechanical element must possess a \textit{sufficiently homogeneous mode}. This means that the mechanical element must be allowed to oscillate back and forth,
when homogeneously illuminated, without loosing its flatness. In \cite{Ruiz15} we suggested a way for implementing such condition through a quasi-one-dimensional membrane, that is, a membrane clamped by a large aspect-ratio frame, such that only one or a few transverse modes could be excited along the short direction. Under these conditions, the system naturally develops a one-dimensional homogeneous mode along the long direction, which becomes unstable through pattern forming instabilities under appropriate parameter settings. One of the limitations of that scheme is that only one-dimensional (1D) transverse-patterns can be generated.

\begin{figure}[t]
\centering
\includegraphics[width=0.6\textwidth]{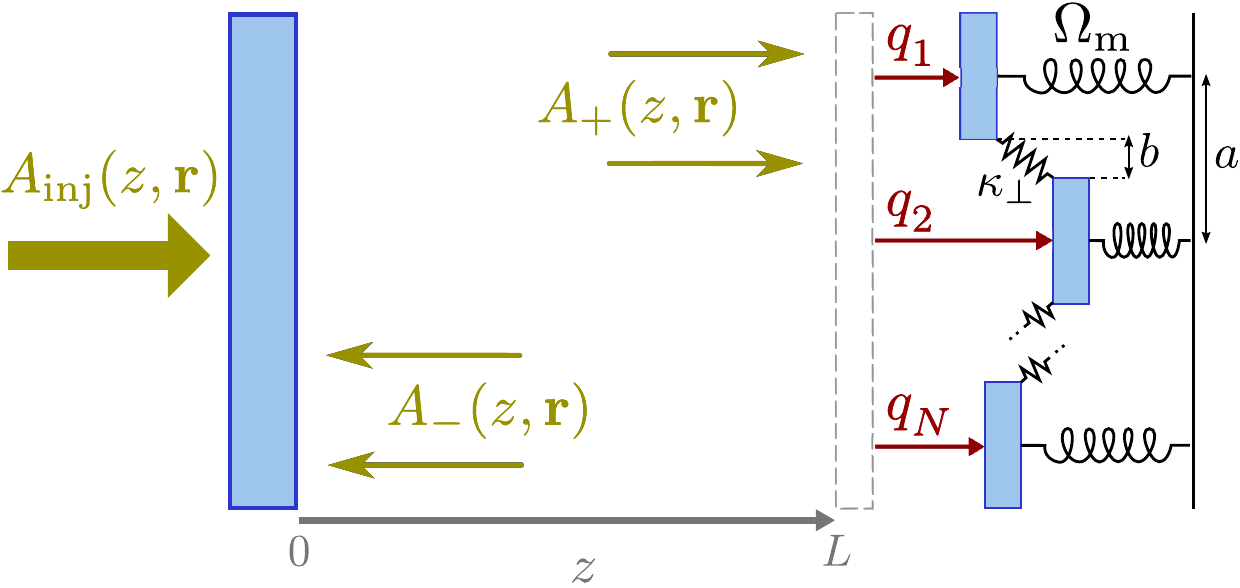}
\caption{Sketch of the OM cavity. The moving mirror is formed by a number of micro-mirrors that are weakly coupled.}
\label{FigScheme}
\end{figure}

In the present paper we investigate a conceptually different possibility of pattern formation to that of \cite{Ruiz15} 
by considering an OM cavity with an oscillating micro-structured end mirror consisting of an array of $N$ 
weakly-coupled micro-mirrors, see Fig. \ref{FigScheme} for a sketch of the 1D case; of course, the system can also be implemented in 2D. This configuration, indeed, allows automatically a homogeneous mode. Below we show that in the limit of large $N$, one recovers the continuous linear-coupling model in~\cite{Ruiz15}, and hence all the predictions in that work apply when the number of elements in the array is large. The interesting point is that one can consider also the limit of small $N$ and study the transition from the continuous to the discrete limit. Among the interesting numerical results that we show below, it is remarkable the fact that a discrete analog
of the continuous-limit cavity solitons can be observed with a relatively small number of coupled micro-mirrors, say $N\approx 10$.

Of course, the problem we are addressing is closely connected with the theory of OM arrays~\cite{Heinrich11,Holmes12,Ludwig2013,Cernotik18} as our model can be thought of as a special implementation of these systems in which all the OM elements are nonlinearly driven through the same multimode intracavity optical field. As the continuum optomechanics theory is recovered from the continuum limit of
the OM array model, this opens the way for a formal connection between pattern formation in OM array cavities and continuum optomechanics.

After this Introduction, we present our model for the OM cavity with micrestructured end-mirror in Section II. 
In Section III we establish the connection with our continuous model of~\cite{Ruiz15} and in Section IV we present some numerical results showing, in particular, the transition from the quasi-continuum limit to a small number of mechanical elements. 
Then, in Section V, we establish a formal connection between the OM array models \cite{Ludwig2013} and the continuum optomechanics, 
as well as we show some relevant numerical results. Finally, in Section VI we give our main conclusions.

\section{Model}

Consider an optical cavity with large-area mirrors, one of which is plane, partially transmitting, and immune to radiation pressure because of its stiffness and mass, while the other is a perfectly reflecting array of weakly coupled micro-mirrors 
(see Fig. \ref{FigScheme}). The field injected in the cavity through the coupling mirror is assumed to be a paraxial, coherent beam
\begin{equation}
E_{\mathrm{inj}}\left( z,\mathbf{r},t\right) =\mathrm{i}\mathcal{V}
A_{\mathrm{inj}}\left( z,\mathbf{r},t\right) e^{\mathrm{i}\left( k_{\mathrm{L}}z-\omega _{\mathrm{L}}t\right) }+\mathrm{c.c.,}
\end{equation}
where $\mathbf{r}=\left( x,y\right) $ denotes the position in the plane transverse to the cavity axis ($z$-axis), and $\mathcal{V}$ is a constant having the dimensions of voltage, which we choose as $\mathcal{V}=\sqrt{\hbar \omega _{\mathrm{c}}/4\varepsilon _{0}L}$ in order to make contact with quantum optics (see Appendix A), $\omega _{\mathrm{c}}$ being the frequency of the longitudinal cavity mode closest to the injected frequency $\omega _{\mathrm{L}}$, with corresponding wave vector $k_{\mathrm{L}}=\omega _{\mathrm{L}}/c$.

The generic intracavity field $E\left( z,\mathbf{r},t\right) $ can be written as
\begin{equation}
E\left( z,\mathbf{r},t\right) =\mathrm{i}\mathcal{V}\left( A_{+}e^{\mathrm{i}k_{\mathrm{L}}z}+A_{-}e^{-\mathrm{i}k_{\mathrm{L}}z}\right) e^{-i\omega _{\mathrm{L}}t}+\mathrm{c.c.,}
\end{equation}
which is the superposition of two waves with slowly varying complex amplitudes $A_{\pm }\left( z,\mathbf{r},t\right) $, propagating along the positive ($A_{+}$) and negative ($A_{-}$) $z$ direction. With similar assumptions as those in~\cite{hyperbolic,Ruiz15} (see Appendix B for the derivation), the field $A_{+}\left( z=L,\mathbf{r},t\right)$ at the microstructured mirror's surface, 
which we denote by $A\left( \mathbf{r},t\right) $, has the following evolution equation 
\begin{equation}
\partial _{t}A=\gamma _{\mathrm{c}}\left( -1+\mathrm{i}\Delta +\mathrm{i}
l_{\mathrm{c}}^{2}\nabla _{\bot }^{2}+\mathrm{i}\frac{4k_{\mathrm{L}}}{T}Q\right) A+\gamma _{\mathrm{c}}\mathcal{E}.  
\label{dAdtaux}
\end{equation}
Here $T$ is the transmissivity of the fixed mirror, $\gamma _{\mathrm{c}}=cT/4L\ $the cavity damping rate, 
$\Delta =\left( \omega _{\mathrm{L}}-\omega _{\mathrm{c}}\right) /\gamma _{\mathrm{c}}$ the dimensionless
detuning parameter, $l_{\mathrm{c}}^{2}=2L/k_{\mathrm{L}}T$ the square of the diffraction length, 
$\nabla _{\bot }^{2}=\partial _{x}^{2}+\partial_{y}^{2}$ the transverse Laplacian, and 
$\mathcal{E}\left( \mathbf{r},t\right) =2T^{-1/2}A_{\mathrm{inj}}\left( L,\mathbf{r},t+t_{\mathrm{c}}\right) $ 
a scaled version of the injection field amplitude ($t_{\mathrm{c}}=2L/c$ is the cavity round-trip time). 
In \eqref{dAdtaux} we have introduced a field $Q\left( \mathbf{r},t\right) $ that measures the local
displacement of the microstuctured flexible mirror perpendicular to its flat state (hence $Q=0$ at rest) 
and next we derive its equation of motion.

We describe the displacement of the microstructured flexible mirror in terms of the individual displacements 
$\{q_{\mathbf{j}}\}_{\mathbf{j\in\mathbb{N}}^{2}}$ of its constituent micro-mirrors, labelled by a double index 
$\mathbf{j}=(j_{x},j_{y})$ in a 2D configuration, as 
\begin{equation}
Q\left( \mathbf{r},t\right) =\sum_{\mathbf{j}}q_{\mathbf{j}}(t)w_{\mathbf{j}}(\mathbf{r}),\label{Q}
\end{equation}
where $w_{\mathbf{j}}(\mathbf{r})$ is a function which equals $1$ when $\mathbf{r}$ is on the surface of micro-mirror $\mathbf{j}$ and is zero otherwise. In the following we assume for simplicity that the micro-mirrors are much larger than the separation between them, that is, $a\gg b$ in Fig. \ref{FigScheme}. Each of these displacements is assumed to satisfy the equation of motion of a damped and forced harmonic oscillator, the force acting on mirror $\mathbf{j}$ having two contributions, 
$F_{\mathbf{j}}=F_{\mathbf{j}}^{(\mathrm{RP})}+F_{\mathbf{j}}^{(\bot )}$, respectively coming from radiation pressure and from the coupling to neighbouring mirrors. The first contribution is readily obtained by integrating the radiation pressure (\ref{RP}) over the surface $\mathcal{S}_{\mathbf{j}}$ of the corresponding micro-mirror
\begin{equation}
F_{\mathbf{j}}^{(\mathrm{RP})}=\frac{2\hbar k_{\mathrm{c}}}{t_{\mathrm{c}}}
\int_{\mathcal{S}_{\mathbf{j}}}d^{2}\mathbf{r}\left\vert A\left( 
\mathbf{r},t\right) \right\vert ^{2},
\end{equation}
where $k_{\mathrm{c}}=\omega _{\mathrm{c}}/c$. As for the force coming from the coupling to neighbouring mirrors, we assume that it originates from a potential that harmonically couples neighbours as 
$V_{\mathbf{j}}^{\bot}=\kappa _{\bot }\sum_{\langle \mathbf{l\rangle }_{\mathbf{j}}}(q_{\mathbf{l}}-q_{\mathbf{j}})^{2}/2$, 
where $\langle \mathbf{l\rangle }_{\mathbf{j}}$ means that the sum is performed over nearest neighbours, hence the
corresponding force is computed as $F_{\mathbf{j}}^{(\bot )}=-\partial V_{\mathbf{j}}^{\bot }/\partial q_{\mathbf{j}}$, which reads
\begin{equation}
F_{\mathbf{j}}^{\bot }=\kappa _{\bot }\mathcal{L}_z[q_{\mathbf{j}}],  
\label{juerza}
\end{equation}
where
\begin{equation}
\mathcal{L}_z[\psi_{\mathbf{j}}]\equiv
\sum_{\langle \mathbf{l\rangle }_{\mathbf{j}}}(\psi_{\mathbf{l}}-\psi_{\mathbf{j}}),  
\label{Lz}
\end{equation}
being $\{\psi_\mathbf{j}\}$ any array and $z$ the coordination number. For instance, for an inner point in a two-dimensional square lattice ($z=4$), 
\begin{equation}
\mathcal{L}_z[\psi_{\mathbf{j}}]=  \psi_{\mathbf{j}+\mathbf{x}}
+\psi_{\mathbf{j}-\mathbf{x}}+
\psi_{\mathbf{j}+\mathbf{y}}+
\psi_{\mathbf{j}-\mathbf{y}}-z\,
\psi_{\mathbf{j}},
\end{equation}
where $\mathbf{x}$ and $\mathbf{y}$ are unit vectors along the $x$ and $y$ axes, respectively.
For points at the boundary trivial modifications can be done to this expression. Note that $\mathcal{L}_z[\psi_{\mathbf{j}}]$ is the finite-difference approximation to the Laplacian of a continuous field $\Psi$ of which $\{\psi_\mathbf{j}\}$ is a sampling. 

Putting everything together we get
\begin{equation}
\label{oscillator}
\ddot{q}_{\mathbf{j}}+\gamma _{\mathrm{m}}\dot{q}_{\mathbf{j}}
+\Omega _{\mathrm{m}}^{2}q_{\mathbf{j}}-\frac{\kappa _{\bot }}{m}\mathcal{L}_z[q_{\mathbf{j}}]=
\frac{2\hbar k_{\mathrm{c}}}{t_{\mathrm{c}}m}\int_{\mathcal{S}_{\mathbf{j}}}d^{2}\mathbf{r}\left\vert A\left( \mathbf{r},t\right) \right\vert^{2},
\end{equation}
with $\gamma _{\mathrm{m}}$, $\Omega _{\mathrm{m}}$ and $m$ the damping rate, oscillation frequency, and mass of the micro-mirrors, respectively. \eqref{oscillator} together with the optical field \eqref{dAdtaux} and definition \eqref{Q} form the equations of our model.

\section{Continuous limit}

Now we consider the limit in which the number of micro-mirrors per unit
length tends to infinity while keeping a finite mass density and sound speed. We first write
the displacements as a function of the continuous mechanical field $Q$ as%
\begin{equation}
q_{\mathbf{j}}=\int_{\mathbb{R}^{2}}\frac{d^{2}\mathbf{r}}{a^{2}}Q\left( \mathbf{r}\right) w_{\mathbf{j}}(%
\mathbf{r}).
\end{equation}
Next, using the immediate properties 
\begin{subequations}
\begin{eqnarray}
\hspace{-10mm}\int_{\mathcal{S}_{\mathbf{j}}}d^{2}\mathbf{r}\left\vert A\left( 
\mathbf{r},t\right) \right\vert ^{2} &=&\int_{\mathbb{R}^{2}}d^{2}\mathbf{r}\left\vert A\left( \mathbf{r},t\right)
\right\vert ^{2}w_{\mathbf{j}}(\mathbf{r}),
\\
\hspace{-10mm}\int_{\mathbb{R}^{2}}d^{2}\mathbf{r}Q\left( \mathbf{r}\right) w_{\mathbf{j}\pm \mathbf{u}}(%
\mathbf{r}) &=&\int_{\mathbb{R}^{2}}d^{2}\mathbf{r}Q\left( \mathbf{r}\mp a\mathbf{u}\right) w_{\mathbf{j}}(%
\mathbf{r}),
\end{eqnarray}
\end{subequations}
with $\mathbf{u}=\mathbf{x},\mathbf{y}$, one gets from the equation of
motion of a generic displacement $q_{\mathbf{j}}$, \eqref{oscillator}, 
\begin{gather}
\partial _{t}^{2}Q\left( \mathbf{r}\right) +\gamma _{\mathrm{m}}\partial
_{t}Q\left( \mathbf{r}\right) +\Omega _{\mathrm{m}}^{2}Q\left( \mathbf{r}%
\right) =\frac{2\hbar k_{\mathrm{c}}a^{2}}{t_{\mathrm{c}}m}\left\vert
A\left( \mathbf{r}\right) \right\vert ^{2}+\notag \\
+\Omega _{\bot }^{2}[Q\left( \mathbf{r}+a\mathbf{x}\right) +Q\left( \mathbf{r%
}-a\mathbf{x}\right)+
+Q\left( \mathbf{r}+a\mathbf{y}\right) +Q\left( \mathbf{r}-a\mathbf{y}%
\right) -4Q\left( \mathbf{r}\right) ],  \label{aux}
\end{gather}%
with $\Omega _{\bot }=\sqrt{\kappa _{\bot }/m}$. Note that we are
considering a micro-mirror that is not at the boundary of the flexible
mirror (as in the continuous limit the
fields extend up to infinity), and hence have used \eqref{juerza}.

The last step consists in taking the limit $a\rightarrow 0$, but keeping
finite both the speed at which transverse perturbations propagate in the
flexible mirror $v=a\Omega _{\bot }$ and its surface mass density $\sigma
=m/a^{2}$. In this limit we can approximate%
\begin{equation}
Q\left( \mathbf{r}\pm a\mathbf{u}\right) \simeq Q\left( \mathbf{r}\right)
\pm a\partial _{u}Q\left( \mathbf{r}\right) +a^{2}\partial _{u}^{2}Q\left( 
\mathbf{r}\right) /2,
\end{equation}
which when used in (\ref{aux}) leads to

\begin{equation}
\partial _{t}^{2}Q+\gamma _{\mathrm{m}}\partial _{t}Q+\left( \Omega _{%
\mathrm{m}}^{2}-v^{2}\nabla _{\bot }^{2}\right) Q=\frac{2\hbar k_{\mathrm{c}}%
}{t_{\mathrm{c}}\sigma }\left\vert A\right\vert ^{2}.  \label{dQ}
\end{equation}%
This is the same equation of motion we derived in the linear--coupling model
of~\cite{Ruiz15}. For sake of later use, we note that in that work it was evidenced the importance of the rigidity parameter $\rho = {v}/{\Omega_\mathrm{m}l_c}$, which can be expressed in terms of our model parameters as
\begin{equation}
    \rho = \frac{a\sqrt{\kappa_\bot/m}}{\Omega_\mathrm{m}l_c}.
\end{equation}

\section{Numerical simulations}
\begin{figure*}[t!]
\flushright\includegraphics[width=0.75\textwidth]{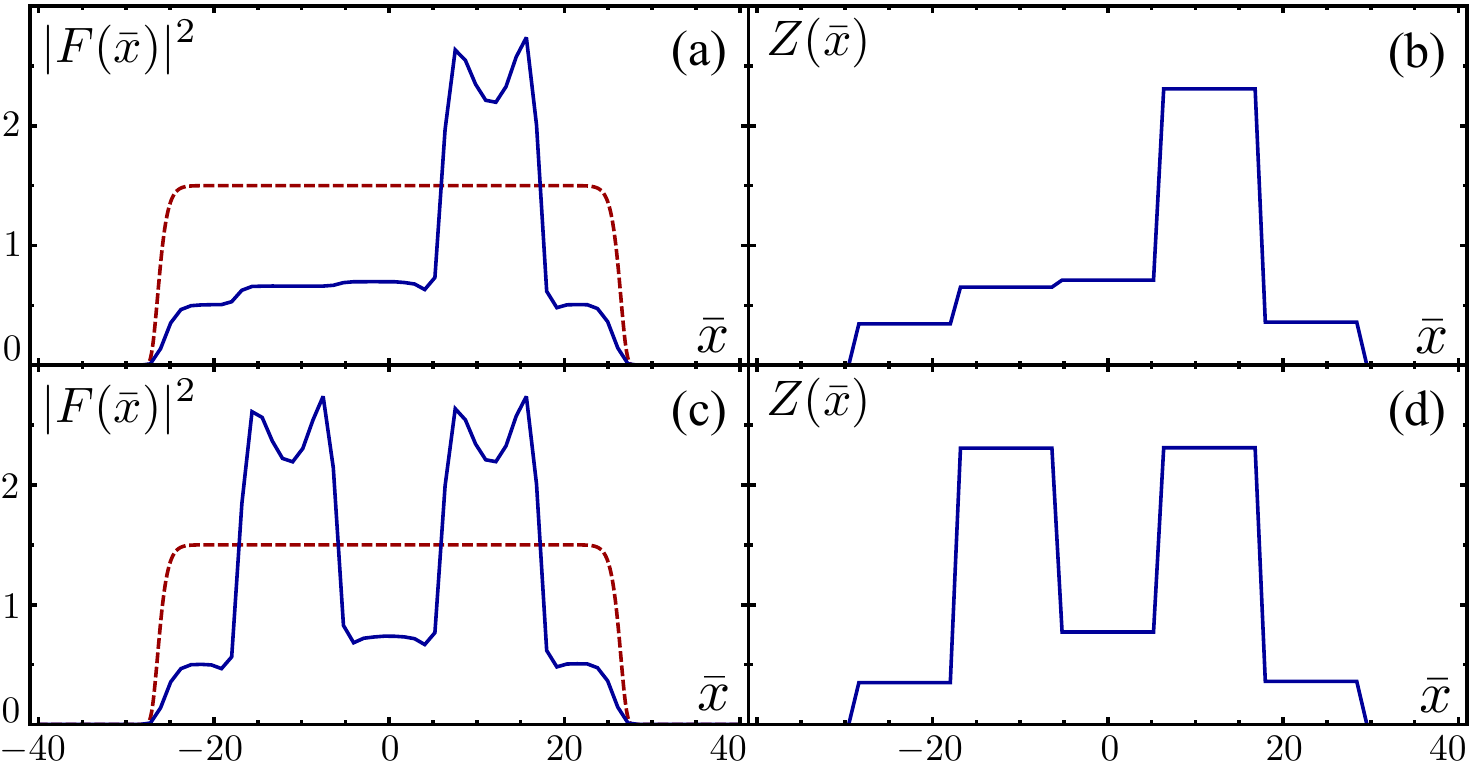}
\caption{Normalized field amplitude (squared) $|F(x)|^{2}$ and mechanical
field $Q(x)$ in the steady-state as a function of the position, for a 1D
system of finite size $x\in [-40l_{\mathrm{c}},40l_{\mathrm{c}}]$. Figs. (a) and (b) show a localized structure that has been written around position $\bar{x}=12$; an additional localized structure has been written around $\bar{x}=-12$ in Figs. (c) and (d), which clearly does not disturb the previous structure. The
basic shape of the injection ($E^{2}$) is represented in dashed red, on top of which a Gaussian profile with the proper width and position is initially fed in order to write the localized structures. The parameters in this simulation are $N=7$, $M=11$, $\protect\gamma =0.1$, $\Omega =10$, $\Delta =-2.2$, $\protect\rho =1.13$, $E_0=\sqrt{1.5}$, and $\sigma_x=23$.}
\label{FigDvsC2}
\end{figure*}

From the previous derivation of the continuous model we can conclude that, at least in the limit of a very large number of micro-mirrors, the system we are proposing is equivalent to that studied in~\cite{Ruiz15}. This implies that all the results that we obtained in~\cite{Ruiz15} for the linear coupling model apply in this limit, both the analytical (concerning homogeneous steady states and their stability properties) and the numerical ones (types of patterns, generalized bistability, temporal dynamics, etc.). Of course, one must wonder how large must the density of micro-mirrors be for the results of the continuous model to still apply, as well as how do the results change when departing from such continuous limit.
\begin{figure*}[t!]
\flushright
\includegraphics[width=0.8\textwidth]{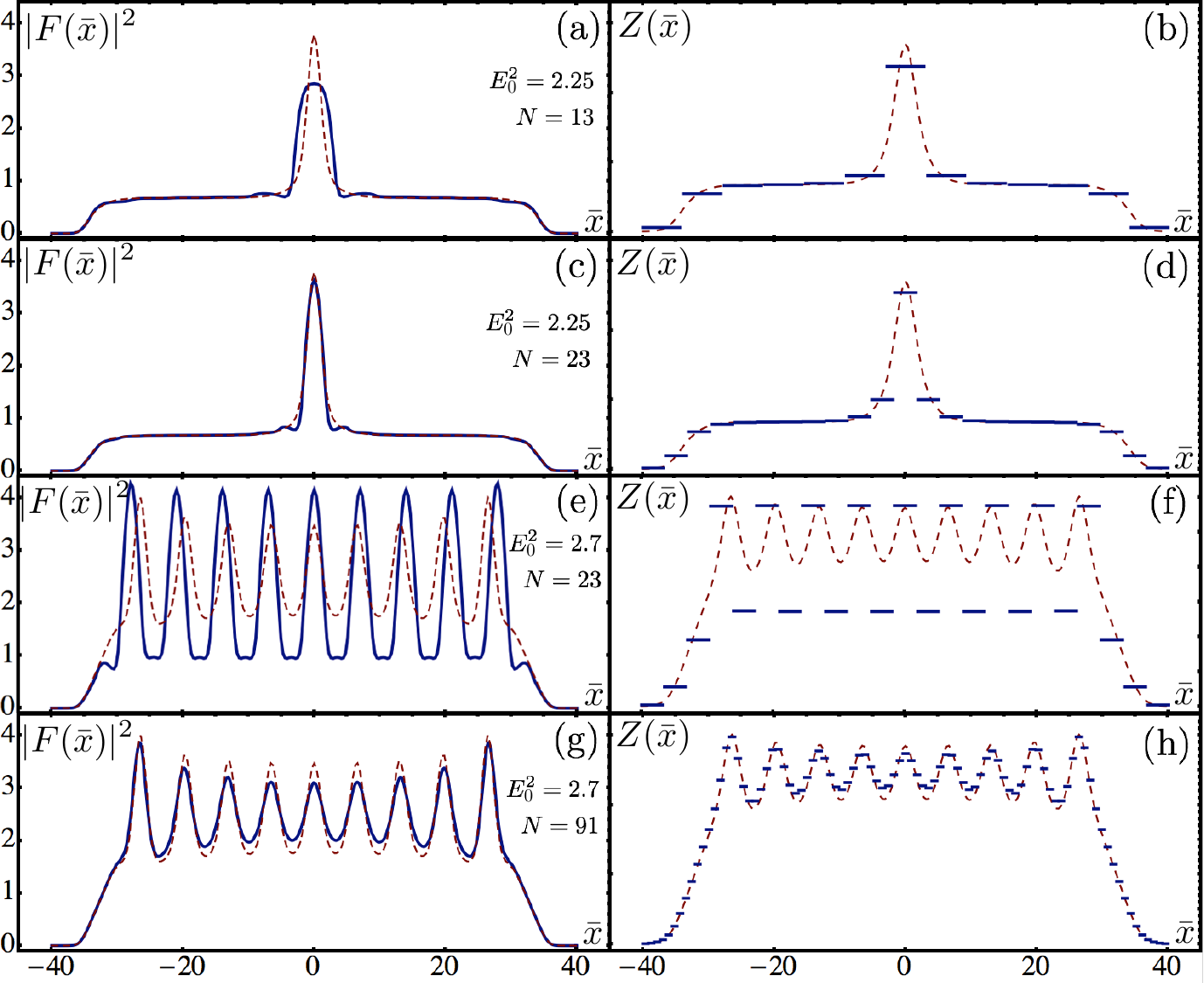}
\caption{Normalized field amplitude (squared) $|F(x)|^{2}$ and mechanical
field $Q(x)$ in the steady-state as a function of the position, for a 1D
system of finite size $x\in \lbrack -40l_{\mathrm{c}},40l_{\mathrm{c}}]$.
The solutions have been obtained by numerical resolution of the discrete
(solid blue) and continuous (dashed red) models described in the text, and
under a top-hat illumination. In all cases $M=11$, $\protect\gamma =0.1$, $\Omega
=10 $, $\Delta =-2.2$, and $\protect\rho =1.13$. The injection values $E_0^{2}$
have been chosen in the region where solitons (a-d) or periodic patterns
(e-h) are expected from the continuous model. For the discrete model we
consider $N$ micro-mirrors as specified in the figure, so that their individual size is 
$a=80l_{\mathrm{c}}/N$.}
\label{FigDvsC}
\end{figure*}
We have performed extensive numerical simulations of both the discrete and continuous models, and Figs. \ref{FigDvsC} 
and \ref{FigDvsC2} summarize our main findings. For simplicity, we have restricted the simulations to one dimension, but similar conclusions are drawn in 2D. We have numerically simulated the continuous model by using the usual split-step method, 
which at any time step provides an approximation of the fields at certain space points. The same method can be applied to the discrete model, and in particular, we take $M$ spatial points for the optical field at every micro-mirror, denoting by $(j,l)$ point $l$ of mirror $j$, so that the field amplitude $A(x)$ is represented by the array $\{A_{j,l}\}_{j=1,2,...,N}^{l=1,2,...,M}$, 
giving a total of $N\times M$ points. The next step consists in choosing a finite-differences form of the integral appearing in the mechanical equations (\ref{oscillator}). We have found that, for stability purposes, an integration rule of the type
\begin{equation}\label{DiscreteIntegral}
\int_{\mathcal{S}_{j}}dx\left\vert A\left( x,t\right) \right\vert
^{2}\approx a\sum_{l=0}^{M+1}d_{l}\left\vert A_{j,l}\left( t\right)
\right\vert ^{2},
\end{equation}
where $A_{j,0}=A_{j-1,M}$ and $A_{j,M+1}=A_{j+1,1}$, is what works best, that is, we use a discrete representation of the integral over mirror $j$ that includes the last point of the previous mirror and the first point of the next one. The weights satisfy the constrain 
$\sum_{j=0}^{M+1}d_{l}=1$, and we have chosen a second order integration rule $\{d_{l}\}_{l=0,1,...,M+1}=\{1,23,24,24,...,24,23,1\}/24M$ which seems to provide very good convergence properties.

Following our previous work~\cite{Ruiz15} and for the sake of convenience we define the following dimensionless versions of the mechanical displacement and optical field,
\begin{equation}
Z=\frac{4k_{\mathrm{L}}}{T}Q,\ \ \ F=\frac{2}{\Omega _{\mathrm{m}}}\sqrt{\frac{2\hbar k_{\mathrm{c}}k_{\mathrm{L}}a}{t_{\mathrm{c}}mT}}A,
\end{equation}
where $Q$ is defined in \eqref{Q}. We also define the dimensionless injection 
$E=(2/\Omega_\mathrm{m})({2\hbar k_\mathrm{c}k_\mathrm{L}a/t_\mathrm{c}mT})^{1/2}\mathcal{E}$ and write $\kappa _{\bot }/m=v^{2}/a^{2}$ in terms of the effective rigidity parameter $\rho =v/\Omega _{\mathrm{m}}l_{\mathrm{c}}$ which together with the detuning was shown to control the appearance of dissipative structures in the continuous model~\cite{Ruiz15}. We have also introduced normalized versions of other parameters, namely $\gamma=\gamma _{\mathrm{m}}/\gamma _{\mathrm{c}}$, $\Omega =\Omega_{\mathrm{m}}/\gamma _{\mathrm{m}}$.

Combining this normalization with the discrete form (\ref{DiscreteIntegral}) of the integral, and introducing dimensionless versions of time and space, $\tau=\gamma_\mathrm{c}t$ and $\bar{x}=x/l_\mathrm{c}$, respectively, we obtain the normalized equations
\begin{subequations}
\begin{eqnarray}
&&\frac{d^{2}z_{j}}{d\tau ^{2}}+\gamma \frac{dz_{j}}{d\tau }+\Omega ^{2}z_{j}=
\rho ^{2}\Omega ^{2}\frac{l_{\mathrm{c}}^{2}}{a^{2}}\mathcal{L}_2[z_{j}]
+\Omega^{2}\sum_{l=0}^{M+1}d_{l}\left\vert F_{j,l}\right\vert ^{2},
\\
&&\partial _{\tau }F=\left( -1+\mathrm{i}\Delta +\mathrm{i}\partial_{\bar{x}}^{2}+\mathrm{i}Z\right) F+E,  
\label{dF}
\end{eqnarray}
\end{subequations}
with the normalized mechanical field written as $Z(\bar{x})=\sum_{j}z_{j}w_{j}(l_{\mathrm{c}}\bar{x})$ and we remind that 
$\mathcal{L}_2[z_{j}]$ refers to the finite-difference version of the Laplacian, see~\eqref{Lz}.
In order to simulate real experimental conditions, we have assumed a top-hat injection profile with finite width, modelled as a super-Gaussian $E(\bar{x})=E_{0}\exp (-\bar{x}^{20}/2\sigma_x^{20})$. In Fig. \ref{FigDvsC}, we show in solid blue the stationary structures we have found in a spatial window $x\in \lbrack -40l_{\mathrm{c}},+40l_{\mathrm{c}}]$ for different number of micro-mirrors $N$ (whose individual size is then $a=80l_{\mathrm{c}}/N$), taking $M=11$ field points per micro-mirror. We are showing results for the special case $\gamma=0.1$, $\Omega=10$, $\Delta=-2.2$, and $\rho =1.13$, and studied the spatial structures for two values of the injection, 
$E_0^{2}=2.25$ and $2.7$ (with $\sigma_x=40$). For these parameters the continuous limit predicts the appearance of, respectively, cavity solitons and periodic patterns~\cite{Ruiz15}. In the figure we show in dashed red the corresponding structures found in the continuous limit for these same parameters. We see that cavity solitons are better captured with a small number of micro-mirrors than periodic patterns. This is so because we need enough space to hold such an extended structure, but they can still be observed with not so many micro-mirrors, as we show below.

In Fig. \ref{FigDvsC2} we illustrate what happens when the number of micro-mirrors is small, a limit that is easier to implement experimentally. We take $N=7$ micro-mirrors and show how localized structures can be supported by a single micro-mirror. We are able to write and erase such structures by an additional Gaussian optical injection at the desired position, and hence they are equivalent to the cavity solitons present in the continuous case.
We find it remarkable that with such a small number of micro-mirrors the cavity solitons of the continuous model can still be recovered.

Once we have demonstrated that the predictions of the continuous model are robust for the discrete model, we pass to study a factor that may severely affect pattern formation, namely boundary effects.
When a homogeneous solution is stable, any perturbation of it will be damped and will tend to disappear. However, the decay of the perturbation may be accompanied by spatial oscillations with characteristic spatial decay and a wavenumber~\cite{SanchezMorcillo1999}, so that depending on the relation between these quantities and the extension of the array and the individual micro-mirror size, resonances may occur. As the boundaries of the mechanical array constitute a perturbation of the homogeneous solution, we can expect a strong influence of their presence in the patterns developed by the system.
\begin{figure}[t!]
\flushright
\includegraphics[width=1.\textwidth]{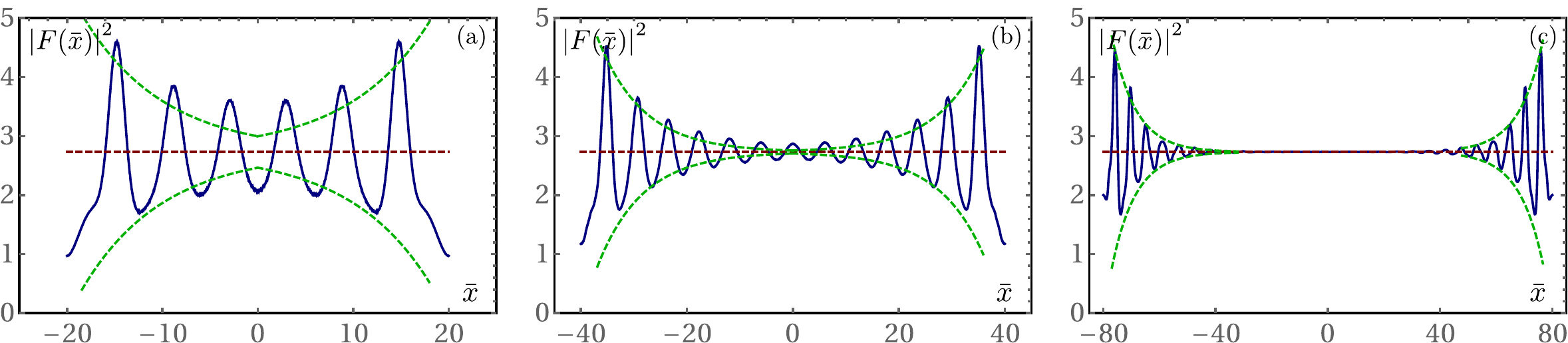}
\caption{Normalized field intensity $|F(x)|^{2}$ (solid blue) in the steady-state as a function of the position, for a 1D system of finite size with: (a) $x\in [-20l_{\mathrm{c}},20l_{\mathrm{c}}]$, (b) $x\in [-40l_{\mathrm{c}},40l_{\mathrm{c}}]$ and (c) $x\in [-80l_{\mathrm{c}},80l_{\mathrm{c}}]$. The parameters in this simulation are $N=91$, $M=11$, $\protect\gamma =0.1$, $\Omega =10$, $\Delta =-2.2$, $\protect\rho =1.13$ and $E_0=\sqrt{3.5}$. The red horizontal dashed curve represents the homogeneous solution expected from the continuous model, and the green dashed curve is obtained by fitting the envelope of the damped spatial modulation, it corresponds to a relaxation characteristic length of $\lambda_{\mathrm{relax}}\approx8.5l_c$.}
\label{FigRelax}
\end{figure}

We first illustrate the spatial oscillations in Fig. \ref{FigRelax}. For the parameters chosen (the system is driven by $E_0^2=3.5$), the system has a stable homogeneous solution $\bar{Z}=\left|\bar{F}\right|^2=2.73$ for the continuous model~\cite{Ruiz15}, however the numerical simulation shows damped spatial oscillations starting from the two edges. We compare three different situations where the spatial extension of the mirror array is $40 l_c$, $80 l_c$ and $160 l_c$ (see Fig. \ref{FigRelax}(a), (b) and (c) respectively). For all the other parameters fixed, the characteristic decay length is the same for the three situations. In particular by fitting the dashed green curve to the envelope of the damped oscillations we obtain $\lambda_{\mathrm{relax}}\approx8.5l_c$. When the size of the system is comparable to this length we observe the formation of a stable pattern (Fig.~\ref{FigRelax}(a)). On the contrary when the system size becomes larger, then spatial oscillating perturbation is completely damped at the center where the homogeneous solution is re-established (Fig.~\ref{FigRelax}(c)).

\begin{figure}[t!]
\flushright
\includegraphics[width=1.\textwidth]{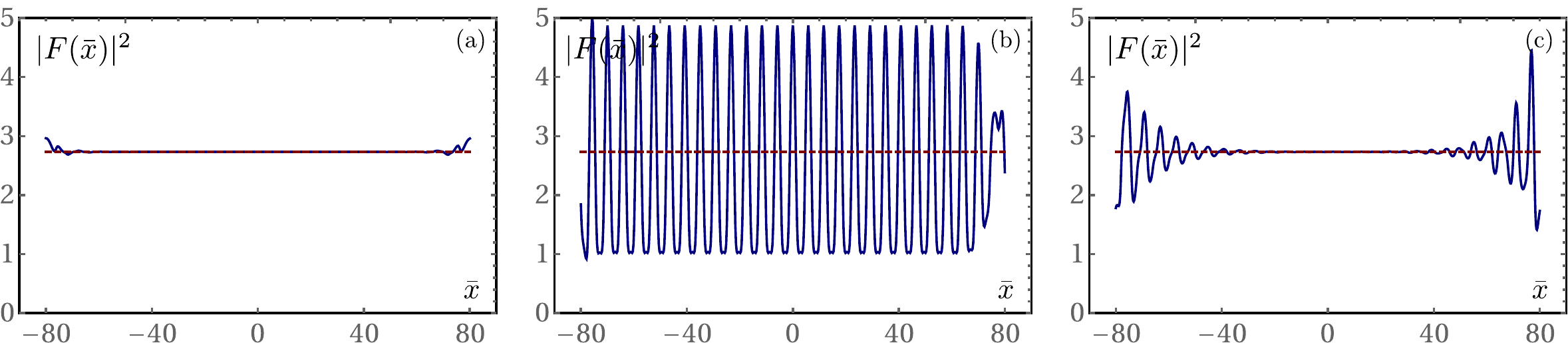}
\caption{Normalized field intensity $|F(x)|^{2}$ (solid blue) in the steady-state as a function of the position, in a 1D system of finite size with $x\in [-80l_{\mathrm{c}},80l_{\mathrm{c}}]$ for (a) $N=40$, (b) $N=55$ and (c) $N=80$. The parameters in this simulation are $M=11$, $\protect\gamma =0.1$, $\Omega =10$, $\Delta =-2.2$, $\protect\rho =1.13$ and $E_0=\sqrt{3.5}$. The red horizontal dashed curve represent the homogeneous solution expected from the continuous model.}
\label{Fig:ResonanceA}
\end{figure}

As stated, because of the discreteness of the mirror array, the perturbation induced by the boundaries gives origin to another effect when it resonates with the spatial period of the mirror array. In this case we observe the emergence of a pattern of spatial period $\lambda_c=2\pi/\mathrm{Re}[k_c]$ when the size $a$ of individual mirror is such that
\begin{equation}
a\approx\lambda_c/2
\label{resonance}
\end{equation}
where $k_c$ is the critical wave vector (in general a complex number) for which one of the eigenvalues of the linear stability analysis is equal to zero. In practical terms, we find it by setting to zero (C12d) in~\cite{Ruiz15}. In Fig.~\ref{Fig:ResonanceA} we present the case of an array of mirrors of total size equal to $160 l_c$, when the system is driven in the upper homogeneous branch by a pump field $E_0^2=3.5$, for three different values of mirror size. In Fig.~\ref{Fig:ResonanceA}(a) we choose $N=40$ so that $a=4$, 
in Fig.~\ref{Fig:ResonanceA}(b) $N=55$ so that $a\approx2.9$, and in Fig.~\ref{Fig:ResonanceA} (c) $N=80$ so that $a=2$. The critical wavevector for the case considered is such that $\mathrm{Re}[k_c]\approx 1.1$. As discussed in the previous paragraph, the boundary conditions induce a perturbation that is spatially modulated relaxing toward the homogeneous solution with some characteristic length. In the case we are considering there is enough room for these oscillations to relax on the homogeneous solution.
Indeed in Figs.~\ref{Fig:ResonanceA}(a) and (c) the spatial period of the discrete array is very different from 
$\lambda_c/2\approx2.85$, while for (b) expression~\eqref{resonance} is matched so that a periodical pattern emerges.

\section{Connection with the standard OM array theory}

In the limit where the intracavity optical field varies slowly enough with respect to the size $a$ of individual mirror and the diffraction length $l_c$ is larger than $a$, we have about one optical mode locally interacting with one mirror. In this case the model defined by eqs. \eqref{dAdtaux} and \eqref{oscillator} can be mapped onto the system of OM arrays suggested in~\cite{Ludwig2013} where 
``\textit{[...] a localized mechanical mode interacts with one laser-driven cavity} (optical) \textit{mode [...] and where both photons and phonons can hop between neighboring sites}".

Then we can use the mesh $\mathbf{r}_{\mathbf{j}}$ of the 2D space with elemental cell of size $a$ for the optical field and we can approximate the Laplacian in \eqref{dAdtaux} with its finite-differences expression of~\eqref{Lz},
\begin{align}
\nabla^{2}_{\bot}A
&\approx 
\frac{1}{a^2}\mathcal{L}_{z}[A_{\mathbf{j}}].
\end{align}
As a consequence \eqref{dAdtaux} can be mapped onto the classical part of the quantum Langevin equation for the optical field $\alpha_{\mathbf{j}}$ in~\cite{Ludwig2013}
\begin{align}
\frac{d \alpha_{\mathbf{j}}}{dt}
=&
\left[
-\frac{\bar{\kappa}}{2}
+\mathrm{i}\bar{\Delta}
+\mathrm{i}g_0 \bar{q}_{\mathbf{j}}
\right]
\alpha_{\mathbf{{j}}}
+
\mathrm{i}\frac{J}{z}\sum_{\langle\mathbf{l}\rangle_{\mathbf{j}}}
\alpha_{\mathbf{l}}
-
\mathrm{i}\alpha_{L}\label{Amarq}
\end{align}
and \eqref{oscillator} can be mapped onto the classical part of the quantum Langevin equation for the mechanical field $\beta_{\mathbf{j}}$ in~\cite{Ludwig2013}
\begin{subequations}
\begin{align}
\frac{d \bar{q}_{\mathbf{j}}}{dt}&=
\bar{\Omega}\bar{p}_{\mathbf{j}},\label{qmarq}
\\
\frac{d \bar{p}_{\mathbf{j}}}{dt}&=
-\frac{\Gamma}{2}\bar{p}_{\mathbf{j}}
-\bar{\Omega}\bar{q}_{\mathbf{j}}
+2g_0\left|\alpha_{\mathbf{j}}\right|^2
+\frac{K}{z}\sum_{\langle\mathbf{l}\rangle_{\mathbf{j}}}
\bar{q}_{\mathbf{l}}\label{pmarq}
\end{align}\label{mechaeq}
\end{subequations}
where $\bar{q}_{\mathbf{j}}=\beta_{\mathbf{j}}+\beta_{\mathbf{j}}^{*}$, $\bar{p}_{\mathbf{j}}=
-\mathrm{i}\left(\beta_{\mathbf{j}}-\beta_{\mathbf{j}}^{*}\right)$ and
\begin{subequations}
\begin{align}
\alpha_{\mathbf{j}}&=a A_{\mathbf{j}},
\\
\bar{q}_{\mathbf{j}}&=q_{\mathbf{j}}/\Delta q_0,
\\
\bar{p}_{\mathbf{j}}&=\dot{q}_{\mathbf{j}}/\Delta p_0,
\end{align}
\end{subequations}
with $\Delta q_0=\sqrt{\hbar/2 m \bar{\Omega}}$ and $\Delta p_0=\hbar/\Delta q_0$ the zero point fluctuations.
Then the one-to-one correspondence with the dynamics of discrete OM arrays such as in~\cite{Ludwig2013} is established by the following correspondence between the parameters of the two models
\begin{subequations}
\begin{align}
\frac{K}{z}&=\frac{\kappa_{\bot}\Delta q_0}{\Delta p_0},
\\
\frac{J}{z}&=
\frac{\gamma_c l_c^2}{a^2},
\\
\bar{\Delta}&=\gamma_c\left(\Delta-z \frac{l_c^2}{a^2}\right),
\\
\bar{\Omega}&=\sqrt{\Omega_{\mathrm{m}}^2+z \kappa_{\bot}/m}.
\end{align}
\end{subequations}
The above derivation, thus, allows the connection between our model and the standard OM array model~\cite{Ludwig2013} and, given that both discrete models are discrete versions of continuous models, the derivation also establishes the relation between our pattern forming model and the continuum optomechanics theory of~\cite{Ludwig2013}.

Next we proceed to numerically illustrate the type of patterns appearing in this standard OM array limit. On the basis of this correspondence we have chosen an array of $N=32$ mirrors with total size of $32 l_c$ ($x\in[-16l_c,16l_c]$) such that the size of each mirror is $a=l_c$. This means that we are just at the limit where our model can be mapped on the discrete model. For improving the validity of the correspondence between the two models (i.e. $a< l_c$) either one can increase the number of mirrors or reduce the size of the array. While the first option has, in principle, no limits apart from that of having enough computational power or -- experimentally -- to have a large array, the second option is physically limited by the fact that the total size of the array should provide enough room for the emergence of structures that have a characteristic size of the order of $l_c$. For a total array size smaller than $32l_c$ we observed a reduced capacity of sustaining structures. Finally we have chosen the remaining parameters as in the previous section: $\gamma=0.1$, $\Omega=10$, $\Delta=-2.2$ and $\rho=1.13$.
The corresponding parameters of the OM array model, given by eqs. \eqref{Amarq} and \eqref{mechaeq},
are $K\approx0.82$, $J\approx3.67$, $\bar{\Delta}\approx-5.87$ and $\bar{\Omega}\approx23.85$. 

\begin{figure*}[t!]
\flushright
\includegraphics[width=1.0\textwidth]{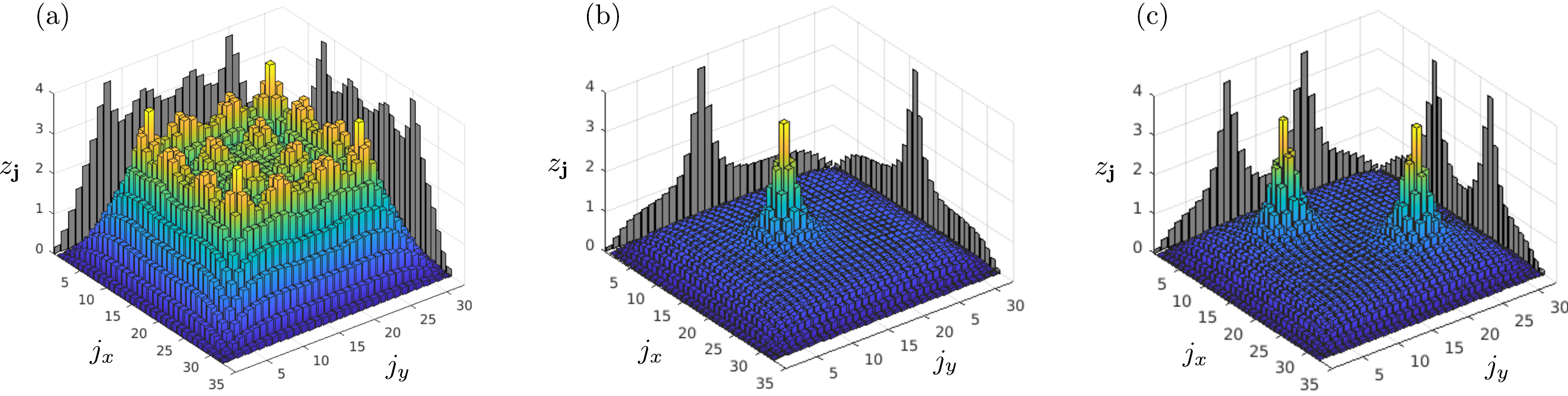}
\caption{Simulations of the standard OM array limit. Normalized mechanical displacement $z_{\mathbf{j}}$ relative to the $\mathbf{j}$-th OM cell (with $\mathbf{j}=(j_x,j_y)$) in the steady state for a $32\times32$ square lattice. Figure (a) shows periodic pattern corresponding to an injection $E_0^2=3$; figure (b) shows a soliton for an injection $E_0^2=2.35$. This solution emerges from instability of the homogeneous solution in the upper branch of the steady state curve. It coexists with steady state homogeneous solution in the lower branch. Figure (c) shows the steady state after the writing of two solitons centered at $\mathbf{j}=(9,9)$ and $\mathbf{j}=(23,23)$ with injection $E_0^2=2.35$.}
\label{Fig:patternMRQ}
\end{figure*}

By virtue of the model correspondence, the OM array presents the same curve of steady-state homogeneous solutions as model of 
Eqs.~\eqref{dAdtaux} and \eqref{oscillator} (also of~\cite{Ruiz15}) that is bistable for the chosen parameters.
When the system is driven by an injection field $E_0^2=3.0$ (with $\sigma_x=0.95\times16 l_c$), the system spontaneously generates a 
$4\times4$ square pattern both in the optical and mechanical fields. In particular in Fig. \ref{Fig:patternMRQ}(a) we traced the normalized displacement $z_{\mathbf{j}}$ of the $\mathbf{j}$-th OM cell. When the injection field is reduced to $E_0^2=2.35$ the optical and mechanical fields spatially localize in a cavity soliton solution at steady state. In Fig. \ref{Fig:patternMRQ}(b) the mechanical field forms a soliton, the width of which involves $5$ elementary cells. This solution coexists with the homogeneous solution of the lower branch of the steady state curve. As a consequence it is possible to write several localized structures for $E_0^2=2.35$. This is shown in \ref{Fig:patternMRQ}(c) where two solitons have been written at $\mathbf{j}=(9,9)$ and $\mathbf{j}=(23,23)$.
Differently from examples shown in Fig. \ref{FigDvsC2}, in the case of OM arrays it is not possible to write a soliton for a too small number of cells (for example $N=7$). Indeed, in the first situation we are in a limit where $l_c<a$ so that structures smaller than one mirror can be sustained by the optical field and therefore the size of the array can be very small. In the case of OM array model we are in the limit of $l_c\geq a$ for which the characteristic size of self-localized structures is larger than one mirror size. For example we see that in Fig. \ref{Fig:patternMRQ} that these structures involve several times the diffraction length $l_c$ ($\approx7$ with the parameters chosen), therefore they cannot be sustained, in the limit of $l_c>a$ by a too small array. In terms of the model 
eqs. \eqref{qmarq} and \eqref{pmarq} this is equivalent to ask $J/z>\gamma_c$.
This limitation could be circumvented by considering a generalization of the model~\cite{Ludwig2013} where each elementary cell is an OM cavity which is degenerate for at least two optical modes.

\section{Conclusions}

In this paper we have proposed an architecture for an OM cavity that allows for the generation of dissipative structures. The device
consists of an OM cavity with an oscillating end-mirror formed by an array of weakly-coupled micro-mirrors. 
This configuration fulfils the basic requirement necessary for OM dissipative structures: the existence of a homogeneous mechanical transverse mode~\cite{Ruiz15}. This proposal offers then an alternative to our previous one~\cite{Ruiz15} that consisted in mounting a flexible mirror on a large aspect ratio frame (with dimensions $L_{x}\gg L_{y}$).

The model we are proposing coincides mathematically with that studied in~\cite{Ruiz15} in the limit of large density of 
micro-mirrors, making all its analytical and numerical results applicable in such limit. We have numerically shown that this is also the case, more qualitatively, when the number of micro-mirrors is not very large, and have found that with a relatively small number of elements there exist solutions reminiscent of the continuous-case cavity solitons. More concretely, we have found that a discrete model consisting of $N\approx 10$ micro-mirrors with size $a\approx 4l_{\mathrm{c}}$ is enough to observe localized structures exactly as predicted by the continuous limit model. Periodic patterns may require a larger number of micro-mirrors depending on their periodicity, but in any case they should still be well captured with a reasonable number of these (say $N<100$). In the second part of the paper we have also connected our model with the discrete optomechanical arrays model recently put forward in~\cite{Ludwig2013}. The connection appears through the discretized version in~\cite{Ludwig2013} meaning that our OM cavity model (with micro-structured mirror) is equivalent to an OM array when the diffraction length is larger that the individual micro-mirror size. Hence, implementing pattern formation in an microsructured OM-cavity is a way of implementing OM arrays and, with more generality, pattern formation is a robust implementation of continuum optomechanics. We hope that our work be useful in the experimental search of dissipative structures in OM devices.

\ack{We thank Chiara Molinelli for useful discussions in the initial brainstorming phase. This work was funded by Spanish Ministerio de Ciencia, Innovaci\'{o}n y Universidades --- Agencia Estatal de Investigaci\'{o}n, and European Union FEDER (projects FIS2014-60715-P and FIS2017-89988-P). CNB acknowledges additional support from a Shanghai talent program and from the Shanghai Municipal Science and Technology Major Project (Grant No. 2019SHZDZX01). GP acknowledges support from Lille University within the framework ``Internationalisation de la recherche 2019 - collaboration bilat\'{e}rales".}

\appendix
\section{Interpretation of the optical field amplitude and radiation pressure}

In the main text we wrote the electric field propagating to the right as%
\begin{equation}
\mathbf{E}_{+}\left( z,\mathbf{r},t\right) =\mathrm{i}\mathbf{x}\mathcal{V}%
A_{+}\left( z,\mathbf{r},t\right) e^{\mathrm{i}k_{\mathrm{L}}z-\mathrm{i}%
\omega _{\mathrm{L}}t}+\mathrm{c.c.,}  \label{Evec}
\end{equation}%
where here we include the polarization of the electric field (defining the $%
x $ direction with corresponding unit vector $\mathbf{x}$), which was
omitted in the main text for simplicity. The aim of this section is to
explain how the choice $\mathcal{V}=\sqrt{\hbar \omega _{\mathrm{c}%
}/4\varepsilon _{0}L}$ characteristic of quantum optics allows us to give a
simple interpretation to the amplitude $A_{+}$ (and similarly for $A_{-}$
and $A_{\mathrm{inj}}$), as well as writing the expression for the radiation
pressure exerted by $\mathbf{E}_{+}$ in terms of this amplitude.

Let us first remind that, within the paraxial approximation, the magnetic
field associated to (\ref{Evec}) can be written as ($\mathbf{y}$ is the unit
vector in the $y$ direction)%
\begin{equation}
\mathbf{B}_{+}\left( z,\mathbf{r},t\right) =\mathrm{i}\mathbf{y}c^{-1}%
\mathcal{V}A_{+}\left( z,\mathbf{r},t\right) e^{\mathrm{i}k_{\mathrm{L}}z-%
\mathrm{i}\omega _{\mathrm{L}}t}+\mathrm{c.c.}.
\end{equation}%
The corresponding Poynting vector is then written as ($\mathbf{z}$ is the
unit vector in the $z$ direction)%
\begin{equation}
\mathbf{S}_{+}=\frac{1}{\mu _{0}}\mathbf{E}_{+}\times \mathbf{B}_{+}=-\frac{%
\mathcal{V}^{2}\mathbf{z}}{\mu _{0}c}\left( A_{+}e^{\mathrm{i}k_{\mathrm{L}%
}z-\mathrm{i}\omega _{\mathrm{L}}t}-\mathrm{c.c.}\right) ^{2},
\end{equation}%
whose magnitude averaged over an optical cycle%
\begin{eqnarray}
\left. \langle S_{+}\rangle \right\vert _{z=L} &=&\frac{2\pi }{\omega _{%
\mathrm{L}}}\int_{t-\pi /\omega _{\mathrm{L}}}^{t+\pi /\omega _{\mathrm{L}%
}}d\tau \left\vert \mathbf{S}_{+}\left( L,\mathbf{r},\tau \right) \right\vert
\simeq \frac{2\mathcal{V}^{2}}{\mu _{0}c}\left\vert A\left( \mathbf{r}%
,t\right) \right\vert ^{2},  \notag
\end{eqnarray}%
provides the instantaneous measurable power impinging point $\mathbf{r}$ of
the mirror located at $z=L$ per unit area (irradiance). Note that we have
made use of the slowly time-varying nature of the amplitude, and remember
that we defined $A\left( \mathbf{r},t\right) =A_{+}\left( L,\mathbf{r}%
,t\right) $ in the main text. Now it is customary in quantum optics to take $%
\mathcal{V}=\sqrt{\hbar \omega _{\mathrm{c}}/4\varepsilon _{0}L}$ so that%
\begin{equation}
\left\vert A\left( \mathbf{r},t\right) \right\vert ^{2}=\frac{t_{\mathrm{c}%
}\left. \langle S_{+}\rangle \right\vert _{z=L}}{\hbar \omega _{\mathrm{c}}},
\end{equation}%
can be interpreted as the number of photons per unit area which arrive to
point $\mathbf{r}$ of the mirror during a round-trip ($t_{\mathrm{c}}=2L/c$
is the cavity round-trip time). With this choice, the theory is quantized by
interpreting this amplitude as an operator satisfying equal-time commutation
relations $[\hat{A}\left( \mathbf{r},t\right) ,\hat{A}^{\dagger }\left( 
\mathbf{r}^{\prime },t\right) ]=\delta (\mathbf{r}-\mathbf{r}^{\prime })$
and $[\hat{A}\left( \mathbf{r},t\right) ,\hat{A}\left( \mathbf{r}^{\prime
},t\right) ]=0$.

From the Poynting vector, we can get the radiation pressure exerted onto a
point $\mathbf{r}$ of the flexible mirror as $P(\mathbf{r},t)=\left. \langle
S_{+}\rangle \right\vert _{z=L}/c$; this is a quantity of fundamental
relevance to our work, and in our case takes the particular expression%
\begin{equation}
P(\mathbf{r},t)=\frac{\hbar k_{\mathrm{c}}}{t_{\mathrm{c}}}\left\vert
A\left( \mathbf{r},t\right) \right\vert ^{2}.  \label{RP}
\end{equation}%
Given our interpretation of $\left\vert A\left( \mathbf{r},t\right)
\right\vert ^{2}$, this coincides precisely with the total momentum
(momentum per photon $\times $ number of photons) hitting point $\mathbf{r}$
of the flexible mirror per unit time and area.

\section{Derivation of the light field equation}

Here we derive equation (\ref{dAdtaux}) of the main text. To this aim we use
the approach of references~\cite{hyperbolic,Ruiz15}, which consists in propagating
the complex amplitudes $A_{\pm }\left( z,\mathbf{r},t\right) $ along a full
cavity round-trip. Assuming that they are slowly varying in space
and time, they satisfy the paraxial wave equation%
\begin{equation}
\left( \partial _{z}\pm c^{-1}\partial _{t}\right) A_{\pm }=\pm \frac{%
\mathrm{i}}{2k_{\mathrm{L}}}\nabla _{\bot }^{2}A_{\pm }.
\end{equation}
Given the amplitude $A_{+}\left( z=L,\mathbf{r},t\right) $, after reflection
on the flexible mirror we get 
\begin{equation}
A_{-}\left( L,\mathbf{r},t\right) e^{-\mathrm{i}k_{\mathrm{L}%
}L}=-A_{+}\left( L,\mathbf{r},t\right) e^{\mathrm{i}k_{\mathrm{L}%
}[L+2Q\left( \mathbf{r},t\right) ]},
\end{equation}
where $Q\left( \mathbf{r},t\right) $ represents the displacement of the
mirror from its rest position ($Q=0$ at rest). The amplitude $A_{-}\left( L,%
\mathbf{r},t\right) $ propagates from $z=L$ to $z=0$ giving rise to a new
amplitude 
\begin{equation}
A_{-}\left( 0,\mathbf{r},t+\frac{1}{2}t_{\mathrm{c}}\right)
=U_{L}A_{-}\left( L,\mathbf{r},t\right) ,
\end{equation}%
where%
\begin{equation}
U_{L}=\exp \left[ \mathrm{i}(L/2k_{\mathrm{L}})\nabla ^{2}\right] ,
\end{equation}%
is the paraxial propagation operator in free space. After reflection onto
the coupling mirror, a new amplitude 
\begin{align}
A_{+}\left( 0,\mathbf{r},t+\frac{1}{2}t_{\mathrm{c}}\right) &=-\sqrt{R}%
A_{-}\left( 0,\mathbf{r},t+\frac{1}{2}t_{\mathrm{c}}\right)
+\sqrt{T}A_{\mathrm{inj}}\left( 0,\mathbf{r},t+\frac{1}{2}t_{\mathrm{c}%
}\right),
\end{align}
is got, with $R$ and $T$ the reflectivity and transmissivity factors of the
coupling mirror, respectively ($R+T=1$ is assumed: lossless mirror).
Finally, propagation from $z=0$ to $z=L$ yields $A_{+}\left( L,\mathbf{r}
,t+t_{\mathrm{c}}\right) =U_{L}A_{+}\left( 0,\mathbf{r},t+\frac{1}{2}t_{
\mathrm{c}}\right) $. Adding all parts together one gets
\begin{align}
A\left( \mathbf{r},t+t_{\mathrm{c}}\right) & =\sqrt{R}e^{2\mathrm{i}k_{
\mathrm{L}}L}U_{L}^{2}\exp \left[ 2\mathrm{i}k_{\mathrm{L}}Q\left( \mathbf{r}
,t\right) \right] A\left( \mathbf{r},t\right)
+\sqrt{T}A_{\mathrm{inj}}\left( L,\mathbf{r},t+t_{\mathrm{c}}\right) ,
\end{align}
where we used $U_{L}A_{\mathrm{inj}}\left( 0,\mathbf{r},t+\tfrac{1}{2}t_{
\mathrm{c}}\right) =A_{\mathrm{inj}}\left( L,\mathbf{r},t+t_{\mathrm{c}
}\right) $. We now take into account that $R\rightarrow 1$ (equivalently, $
T\rightarrow 0$) so that $\sqrt{R}=\sqrt{1-T}\rightarrow 1-T/2$. Next we
assume that light is almost resonant with the cavity, specifically we impose
that $2\left( \omega _{\mathrm{L}}-\omega _{\mathrm{c}}\right) L/c=\delta $
is of order $T$, where $\omega _{\mathrm{c}}$ is the cavity longitudinal
mode frequency (hence $\omega _{\mathrm{c}}=m\pi c/L$, $m\in 
\mathbb{N}
$) closest to $\omega _{\mathrm{L}}$, what allows approximating $\exp (2%
\mathrm{i}k_{\mathrm{L}}L)=\exp (2\mathrm{i}\omega _{\mathrm{L}}L/c)\approx
1+\mathrm{i}\delta $. We assume as well that $k_{\mathrm{L}}Q\left( \mathbf{r%
},t\right) \ $is of order $T$ (the mirror displacement/deformations are much
smaller that the optical wavelength), so that $\exp \left[ 2\mathrm{i}k_{%
\mathrm{L}}Q\left( \mathbf{r},t\right) \right] \approx 1+2\mathrm{i}k_{%
\mathrm{L}}Q\left( \mathbf{r},t\right) $. Similarly we assume that the
effect of diffraction is small (this implies that both mirrors must be
sufficiently close each other, either physically or by means of lenses) so
that we can expand $U_{L}^{2}\approx 1+\mathrm{i}(L/k_{\mathrm{L}})\nabla
_{\bot }^{2}$. All these assumptions imply that the overall variation of $A$
between consecutive round-trips is very small and then one can approximate $%
\partial _{t}A$ by $\left[ A\left( \mathbf{r},t+t_{\mathrm{c}}\right)
-A\left( \mathbf{r},t\right) \right] t_{\mathrm{c}}^{-1}$. With all these
approximations we get, to the lowest nontrivial order,%
\begin{equation}
\partial _{t}A\left( \mathbf{r},t\right) =\gamma _{\mathrm{c}}\left( -1+%
\mathrm{i}\Delta +\mathrm{i}l_{\mathrm{c}}^{2}\nabla _{\bot }^{2}+\mathrm{i}%
\frac{4k_{\mathrm{L}}}{T}Q\right) A+\gamma _{\mathrm{c}}\mathcal{E},
\end{equation}%
where all the parameters are defined in the main text; this is precisely
\eqref{dAdtaux}, and it is the same light-field equation we derived in the
linear-coupling model of~\cite{Ruiz15}.

\vspace{12pt}
\hrule

\end{document}